\documentclass[12pt]{article}
\usepackage{amsmath}
\usepackage{epsf,afterpage,epsfig}

\def\abstracts#1#2#3{{
        \centering{\begin{minipage}{4.62in}\baselineskip=13pt
        \small
        \centerline{\bf Abstract}
        \vspace*{0.2cm}                
        \parindent=0pt #1\par
        \parindent=18pt #2\par
        \parindent=15pt #3
        \end{minipage} }\par}}
\begin{document}
\vspace*{-2cm}
\hfill \parbox{4cm}{ ~\\~}\\
%
\centerline{\LARGE \bf The Wrong Kind of Gravity}\\[0.3cm]
\vspace*{0.2cm}
\centerline{\large {\em Wolfhard Janke$^1$ and 
Desmond A. Johnston$^2$\/}}\\[0.4cm]
\centerline{\large {\small $^1$ Institut f\"ur Theoretische Physik,
                    Universit\"at Leipzig}}
\centerline{    {\small D-04109 Leipzig, Germany }}\\[0.5cm]
\centerline{\large {\small $^2$ Department of Mathematics, 
                    Heriot-Watt University}} 
\centerline{    {\small Edinburgh, EH14\,4AS, Scotland  }}\\[0.5cm]
\abstracts{}{
The KPZ formula \cite{kpz}
shows that coupling central charge $c \le 1$ spin models to 2D quantum
gravity
dresses the conformal weights to get new critical exponents,
where the relation between the original and dressed weights
depends only on $c$.
At the discrete level the coupling to 2D gravity is effected by putting
the spin models on annealed ensembles of $\Phi^3$
planar random graphs or their dual
triangulations, where the connectivity fluctuates on the same
time-scale as the spins.\\
\indent
Since the sole determining
factor in the dressing is the central charge,
one could contemplate putting a
spin model on a quenched
ensemble of 2D gravity graphs
with the ``wrong'' $c$ value.
We might then expect to see the critical exponents
appropriate to the $c$ value used in generating the graphs.
In such cases the KPZ formula
could be interpreted as giving a continuous line of critical exponents
which depend on this central charge.
We note that rational exponents other than the
KPZ values can be generated using this procedure
for the Ising, tricritical Ising and 3-state Potts models.
}{}
%
\thispagestyle{empty}
\newpage
\pagenumbering{arabic}
%
                     \section{Introduction}
%
Among the many puzzles resolved 
by 2D conformal field theory is the appearance
of rational critical exponents
in models such as the 2D Ising and Potts models. The miracle
is repeated when the models are coupled  to 2D quantum gravity
since, as was shown by KPZ \cite{kpz} and DDK \cite{ddk},
the dressing of the conformal weights 
by the Liouville field of 2D
quantum gravity leads to a new set of exponents which are 
nonetheless still rational. The key formula in establishing
the relation between the bare ($\Delta$) and dressed ($\tilde \Delta$)
conformal weights
\begin{equation}
\tilde \Delta = { \sqrt{1 - c + 24 \Delta} - \sqrt{1 -c } \over \sqrt{25 -c} - \sqrt{1 -c} }
\end{equation}
depends only on the central charge $c$ of the matter. The net effect of the 
gravitational
dressing for the minimal $(p,q)$ conformal models with 
$c = 1 - 6 (p-q)^2/pq$, where the primary scaling
operators
are labelled by two integers $r,s$ satisfying
$1 \le r \le p-1$, $1 \le s \le q - 1$, is to transmute the bare weights 
$\Delta_{r,s} = \Delta_{p-r,q-s} = [(rq-sp)^2 - (p-q)^2]/4pq$
from the Kac table
into
\begin{equation} 
\tilde \Delta_{r,s} = { | rq - sp | - | p - q | \over p + q - | p - q| },
\end{equation}   
where $| p - q | = 1$ for unitary models.
The relation between $\Delta$ and $\tilde \Delta$ may be written as
\begin{equation}
\tilde \Delta - \Delta = - {\xi^2 \over 2} \tilde \Delta ( \tilde \Delta - 1),
\label{e00}
\end{equation}
where
\begin{equation}
\xi = - { 1 \over 2 \sqrt{3} } ( \sqrt{ 25 -c } - \sqrt{ 1 -c} ),
\label{e01}
\end{equation}
and is called the KPZ scaling relation.

The effect of coupling various statistical mechanical models to 2D gravity, 
such as the Ising and $q \le 4$ Potts models,
can thus be calculated using the KPZ/DDK results.
If we denote the critical temperature for the 
phase transition in these models by $T_c$ and the reduced temperature $|T - T_c|/T_c$ by $t$ then the
critical exponents $\alpha, \beta $
are defined in the standard manner as $t \rightarrow 0$ by 
\begin{eqnarray}
C_{\rm sing} &\sim& t^{-\alpha}, \nonumber \\  
M &\sim& t^{\beta},
\label{e04}
\end{eqnarray}
where $C_{\rm sing}$ is the singular part of the specific heat and  $M$ is the 
magnetization.
It is then possible to calculate $\alpha$ and $\beta$
using the conformal weights of the energy density operator 
$\Delta_{\epsilon}$ and
spin operator $\Delta_{\sigma}$
in both the dressed and undressed cases,
\begin{eqnarray}
\alpha &=& {1 - 2 \Delta_{\epsilon} \over 1 - \Delta_{\epsilon} }, \nonumber \\
\beta &=&  { \Delta_{\sigma} \over 1 - \Delta_{\epsilon} }.
\label{e05}
\end{eqnarray}
The various scaling relations between the critical exponents then allow
the determination of the complete set of exponents. 

\section{When the wrong $c$ can be right}

The preceding derivation is quite natural when considering the models
in a gravitational context. Since the matter is interacting with gravity,
it is the central charge of the  matter itself which gets fed into the KPZ formula
and returns the new set of rational dressed conformal weights and consequently new set of
rational critical exponents. There are, however, circumstances in which one could conceive
of coupling the conformal matter to graphs with the ``wrong'' 
central charge. The first of these
is when one  considers the matter living on a quenched ensemble of 2D gravity graphs, as
was done in \cite{bhj94,0}. In this case the interaction between the graphs and the matter
is switched off and one is in effect looking at an ensemble with quenched connectivity disorder.
This ensemble displays several interesting effects, including a softening of 
first-order phase
transitions in $q>4$ Potts models to continuous transitions and the possible appearance of
a new set of non-rational (but still algebraic) quenched exponents for $q \le 4$ Potts 
models. In these respects it very much resembles the quenched bond disorder models that have
attracted much attention recently \cite{00} rather than
other quenched connectivity disorder ensembles generated using Poisonnian random lattices
\cite{Janke, javi95b}
which retain the characteristics of their regular lattice counterparts.

The relevant
relation between the quenched dressed weights and bare weights 
is given by the $c=0$ version of the 
KPZ formula
\begin{equation}
\tilde \Delta = { \sqrt{1 + 24 \Delta} - {1 } \over 4 }.
\end{equation}
It has recently been pointed out by Cardy \cite{cardy2} that one should,
in fact, see multi-fractal scaling of local correlators on quenched gravity graphs,
just as with quenched bond disorder. The $n$th power of a correlator 
with weight $\tilde \Delta$ averaged over the disorder
scales not as $n \tilde \Delta$, but rather
\begin{equation}
\tilde \Delta_n = { \sqrt{1 + 24 n \Delta} - {1 } \over 4 }.
\end{equation}

Freed from the bounds of using the ``right'' value of $c$ in the KPZ
formula we can consider the effect of coupling conformal matter
to other backgrounds, whether quenched or annealed. 
In the quenched case one is interested in calculating
the (reduced) free energy
$F = [\ln Z]_{av}$
where $[...]_{av}$ is a quenched average over an ensemble of graphs
characterized by a central charge $c=d$.
Such graphs can be generated by using the adjacency matrix of the graph
$G$ since the fixed area (i.e. fixed number of vertices)
partition function $Z_A$ obtained on integrating out $d$
scalar fields with central charge $d$ is
\begin{equation}
 Z_A \;=\; \sum_{G\in{\cal G}(A)} ({\rm det} \; C_G)^{-d/2} \;,
\end{equation}
where $\cal G (A)$ is the
class of graph being being summed over
and $C_G$ is the adjacency matrix of the  the graph $G$:
\begin{equation}
 C_G \;=\; \left \{
 \begin{array}{ll}
  q_i  & \qquad \text{if $i=j$,} \\
  -n_{ij}  & \qquad \text{if $i$ and $j$ are adjacent,} \\
  0     & \qquad \text{otherwise.}
 \end{array}
 \right .
\end{equation}
Since $d$ is now a parameter, one can in principle use
arbitrary negative or positive values in generating the ensemble of graphs.
In the above $q_i$ is the order of vertex $i$ and $n_{ij}$ is
the number of edges connecting the adjacent vertices $i$ and
$j$, which can be more than one in certain ensembles $\cal G (A)$.

In analytical calculations for quenched ensembles we can use the replica 
trick to (formally) replace the free energy by the $n \rightarrow 0$
limit of an $n$-replicated version of our matter action
\begin{eqnarray}
F &=& [\lim_{n \rightarrow 0} (Z^n - 1)/n ]_{av} \nonumber \\
  &=& \lim_{n \rightarrow 0} ([Z^n]_{av} - 1)/n
\end{eqnarray}
which relates the quenched ensemble to the annealed problem with $n$ copies of 
the matter fields
of interest. The $[...]_{av}$ stands as before for a functional integral 
over surfaces with central charge $d$, now dynamically coupled to
the matter fields. The total central charge 
is thus $c_{total} = d + n c_{matter}$ and in the quenched
$n \rightarrow 0$ limit $c_{total} \rightarrow d$ is the number which appears
in the KPZ formula.
A simulation to this effect has already been carried out in \cite{thor} where 
good numerical agreement was obtained for the measured exponents of the 
Ising model on a quenched ensemble of $d= -5$
graphs and those calculated by substituting $c = d = -5$ in the KPZ formula 
for the dressed
energy and spin weights \footnote{The interpretation put upon the 
results there was that annealed and quenched ensemble of graphs gave the same
results provided the total central charge, $c_{total}$, was the same -
the difference is essentially semantic.} .

Since the central charge of the graphs is decoupled from the matter
in quenched simulations such as that described above,
the KPZ formula on such backgrounds can be thought of as giving
a line of dressed conformal weights, say $\tilde \Delta (d)$,  depending on the central charge
associated with the graphs $d$. If we parameterise $d$ in the customary manner
\begin{equation}
d = 1 - { 6 ( p - q)^2 \over p  q}
\end{equation}
we arrive at the following version of the KPZ formula
\begin{equation}
\tilde \Delta (d) = { \sqrt{p^2 + q^2 - 2 p q ( 1 - 2  \Delta) } - | p - q|   \over p + q - | p - q| }
\end{equation}
for the dressing of weights
in a gravitational background characterised by central charge $d$.
The energy and spin weights derived from this formula would then, via equ.(6),
give a line of critical exponents $\alpha (d)$ and $\beta ( d)$ which depended
on the background central charge $d$.
In annealed simulations the central charge of the (now non-replicated)
matter should be included and $d$ is replaced by $c_{total} = d +  c_{matter}$
in the above considerations.

\section{Rational Points}

On a line of continuously varying critical exponents
the rational values are typically the most interesting,
a prime example being the 8 vertex model on a square lattice 
where the Ising model, amongst others, appears at such a point.
We might thus enquire whether rational points
other than the standard ones (i.e. $d=1/2$ for Ising, $7/10$
for the tri-critical Ising model, $4/5$ for the 3-state Potts model)  exist.
In the case of the Ising model $\Delta_{\epsilon} = 1/2$, $\Delta_{\sigma} = 1 /16$
and there are no other operators in the conformal table apart from the unit operator.
If we want to obtain rational $\tilde \Delta_{\epsilon} (d)$ and $\tilde \Delta_{\sigma}(d)$,
and hence rational $\alpha (d)$ and $\beta (d)$, we see from equ.(13)
that both
$p^2 + q^2$ and $p^2 + q^2 - 7 p q / 4$  must be perfect squares. 
The first condition will be satisfied by the first two members of any Pythagorean triple
$(p,q,m)$ \footnote{Pythagorean triples are three integers satisfying
$p^2 + q^2 = m^2$.}
which can be parameterised in general as $p = u^2 - v^2, q = 2 u v,
m=u^2 + v^2$ with $g.c.d. (u,v)=1$.
Inserting this into the second condition and looking at the possible factorisations
shows that only two triples satisfy both conditions, $(3,4,5)$ and $(5,12,13)$.
The first corresponds to $d=1/2$ and, as it should, returns the standard KPZ weights for the
Ising model coupled to 2D gravity. The second, however, is a background with $d= - 39 / 10$
and gives the weights
\begin{eqnarray}
\tilde \Delta_{\epsilon}  \left(  - { 39 \over 10} \right)     &=&  { 3 \over 5} \nonumber \\
\tilde \Delta_{\sigma}  \left(  - { 39 \over 10} \right)     &=&  { 1 \over 10} 
\end{eqnarray}
which translates to exponents $\alpha ( - 39 /10) = -1/2$, 
$\beta (  - 39 /10) = 1 / 4$.
For comparison we show in Table~1 the exponents $\alpha$, $\beta$ and 
$\gamma$ \footnote{
Derived from the scaling relation $\alpha + 2 \beta + \gamma = 2$.}
for the flat lattice (Onsager exponents),
KPZ ($d=1/2$), quenched ($d=0$) and $d = - 39/10$.

\begin{center}
\begin{tabular}{|c|c|c|c|c|c|c|c|c|c|} \hline
$$& $\alpha$  & $\beta$  & $\gamma$ \\[.05in]
\hline
Onsager& $0$  & $\frac{1}{8}$ & $\frac{7}{4}$ \\[.05in]
\hline
KPZ  $(d=1/2)$& $-1$  & $\frac{1}{2}$ & $2$ \\[.05in]
\hline
Quenched $(d = 0)$& $-0.8685169$  & $0.4167516$ & $2.0350137$  \\[.05in] 
\hline
$d = -{39 \over 10}$& $-{1 \over 2} $  & ${ 1 \over 4}$ & $2$  \\[.05in]
\hline
\end{tabular}
\end{center}
\vspace{.1in}

\centerline{Table 1:  Critical exponents for the Ising model}

\medskip

Remarkably, $\alpha(-39/10) = - 1 / 2$ is the standard KPZ value for the 
{\it three}-state Potts model which has $c=4/5$, but $\beta( -39 / 10) =
1/4$ is half the 3-state Potts model exponent. 
This prompts one to look at the weights of the three-state Potts model
in their own right. In this case one has a much larger conformal grid
of allowed scaling dimensions, twelve of which actually appear
as physical operators. Demanding rationality for all of these
turns out to be too restrictive for all but the KPZ 
values with $d=4/5$. However, if we ask for rationality of only the 
energy and spin operators, which have bare weights 
$\Delta_{\epsilon} = 2/5$ and $\Delta_{\sigma} = 1/15$
repectively,
equ.(13) now shows that
$p^2 + q^2 - (2 / 5 ) p q$ and $p^2 + q^2 - ( 26 / 15) p q$ 
must be perfect squares. We again find two possible solutions,
$d=4/5$ and $d = -3886/1115$. The resulting exponents are tabulated in Table~2
along with the classical (fixed lattice), KPZ and quenched values.

\begin{center}
\begin{tabular}{|c|c|c|c|c|c|c|c|c|c|} \hline
$$& $\alpha$  & $\beta$  & $\gamma$ \\[.05in]
\hline
Fixed& $\frac{1}{3}$  & $\frac{1}{9}$ & $\frac{13}{9}$ \\[.05in]
\hline
KPZ $( d = 4/5)$& $-\frac{1}{2}$  & $\frac{1}{2}$ & $\frac{3}{2}$ \\[.05in]
\hline
Quenched $(d=0)$& $-0.2932676$  & $0.3511286$ & $1.5910104$  \\[.05in]
\hline
$d = -3886/1115$& $-\frac{1}{27}$  & $\frac{6}{27}$ & $\frac{43}{27}$ \\[.05in]
\hline
\end{tabular}
\end{center}
\vspace{.1in}                                          

\centerline{Table 2:  Critical exponents for the 3-state Potts model}
  
\medskip

We have omitted the tricritical Ising model,
which strictly speaking comes between the Ising and three-state
Potts model in any classification, in our discussion. Once again
demanding rationality for the full conformal grid is too
restrictive to give any values of $d$ other than $7/10$, 
but we can still get rational values 
by restricting ourselves to 
rational energy and spin operator weights.
In this case the bare weights are
$\Delta_{\epsilon} = 1 / 10$ and $\Delta_{\sigma}
= 3/80$ and
we find  a whole series of additional solutions
$d = -1449/400$, $-3059/1430$, $-133763/156400$, $\ldots$ as well as
(unlike the Ising and 3-state Potts models)
positive solutions $d=69/70, 44719/81200, \dots$. 

                     \section{Conclusions}
%
In summary, we have seen that treating the central charge in the
KPZ formula as a free parameter admits rational exponent values
other than the standard ones for Ising, tricritical Ising
and three-state Potts models. The numerology is most convincing in the Ising
case, since both the operators in its small conformal grid acquire rational
weights when $d = -39/10$. Although  rational weights can be arranged for
the energy and spin operators at novel values of the central charge 
for the other two, the rest of the conformal grid still acquires irrational 
weights. The three-state Potts model has one other rational value of $d=
- 3886/1115$, and the tricritical Ising many, including positive values.

We have noted that feeding a value of the central charge other 
than that of the matter into the KPZ formula is 
precisely what is required when
the matter/gravity back reaction is switched off, as in quenched
simulations. Given the similarity of spin model behaviour
on such ensembles to those with quenched bond disorder and the existence
of non-perturbative results for the resulting exponents, investigation of 
such models may be useful for illuminating some of the murkier properties
of ferromagnetic systems with quenched disorder. 
Formally exponents
calculated for the ``wrong'' central charge values 
can also apply to annealed
ensembles of graphs (i.e. the connectivity in the graphs is fluctuating
on the same time-scale as the spins, even if the back reaction of the matter
on the graphs is switched off) so long as the appropriate matter central charge is included. 

Finally, it is worth noting that the new rational points all appear to be
numerically accessible, since \cite{thor} investigated $c= -5$
and the central charges for the Ising and three-state Potts rational exponents
are both in the vicinity of $-4$.
It would be interesting to investigate
either numerically or analytically whether the rational points
were simply a numerical accident,
or differed in some manner from the rest of the line of exponents,
as well as the occurrence of such points in other known models.
%
%
\section*{Acknowledgements}
DJ was partially supported by a Royal Society of 
Edinburgh/SOEID Support Research Fellowship. 
The collaborative work of DJ and WJ was funded by ARC grant
313-ARC-XII-98/41.
%
%
%
%

%
\end{document}